\documentclass[fleqn,twoside]{article}
\usepackage{espcrc2}
\usepackage[dvips]{graphicx}
\usepackage{amssymb}
\usepackage{amsmath}
\usepackage{pennames}

\newcommand{\D}{\displaystyle}
\newcommand{\PZ}{\mbox{$\mathrm{Z}$}}
\newcommand{\dr}{\mathrm{d}}
\newcommand{\PLL}{$(\Lambda\Lambda, \: \bar{\Lambda}\bar{\Lambda})$}
\def\PLB#1#2#3{{Phys.~Lett.~B}{\bf #1}\ (#2)\ #3}
\def\EPJC#1#2#3{{Euro.~Phys.~J. C~}{\bf #1}\ (#2)\ #3}

\title{Hadronic final state interactions at ALEPH and OPAL}

\author{V.M. Ghete
        \address{Institute for Experimental Physics,  
        University of Innsbruck, 
        A--6020, Innsbruck, Austria}}

\begin{document}
\pagestyle{empty}

\begin{abstract}
The studies of Fermi--Dirac correlations of \PgL\PgL\ and 
$\bar{\Lambda}\bar{\Lambda}$ pairs in hadronic \PZ\ decays,
Bose--Einstein correlations and colour reconnection 
in \PW-pairs decays performed by the ALEPH collaboration in 
\Pep\Pem\ annihilation at LEP are presented. 
%A depletion of 
%events are observed for the region $Q < 2$~GeV in the \PLL\ 
%system. No colour reconnection effects are observed in 
%\PW-pair decays, and the Monte Carlo model with  
%Bose--Einstein correlations between decay products of 
%different \PW's is disfavoured at $2.7\sigma$. 
The OPAL analysis of Bose--Einstein correlations in \PW-pair decays 
is also discussed.
\end{abstract}

\maketitle

\section{Introduction}

Bose--Einstein (BE) correlations between identical bosons and 
Fermi--Dirac (FD) correlations between identical fermions lead 
to an enhancement or a suppression, respectively,  of the particle 
pairs produced close to each other in phase space. The effect is 
sensitive to the distribution of particle sources in space and 
time.
The strength  of the correlations
can be expressed by the two-particle correlation
function $C(p_1, p_2)=P(p_1, p_2)/P_0(p_1, p_2)$, where
$p_1$ and $p_2$ are the four-momenta of the particles, $P(p_1, p_2)$ 
is the measured differential cross section for the pairs and 
$P_0(p_1, p_2)$ is that of a reference sample, identical to the data 
sample in all aspects except the presence of FD or BE correlations. 
Usually $C(Q)$ is measured, where $Q^2=-(p_1-p_2)^2$. 

For \PW-pairs from $\Pep\Pem \to \PWp\PWm$ at energies in the 
LEP2 range, the distance between \PWp\ and \PWm\ vertices 
is less than 0.1~fm, i.e. less than the typical hadronic 
distance scale of 1~fm. Therefore the fragmentation of \PWp\ 
and  \PWm\ may not be independent. Two phenomena may appear: 
pions from different \PW's may exhibit BE
correlations and pairs of quarks and antiquarks  $\Pq_1\Paq_4$ 
and $\Pq_3\Paq_2$ from the decay of different \PW's can form 
colour strings (colour reconnection). Colour reconnection (CR) 
and BE correlations have opposite effects.  
They may influence the accuracy of the \PW\ mass measurement 
at LEP. 

In this talk, three ALEPH analyses are presented: FD 
correlations of \PgL\PgL\ and $\bar{\Lambda}\bar{\Lambda}$ 
pairs in hadronic \PZ\ decays at LEP1~\cite{aleph-FD}, colour 
reconnection~\cite{aleph-CR} and BE correlations~\cite{aleph-BE} 
in \PW-pairs decays at LEP2. 
The OPAL analysis of BE correlations in \PW-pair 
decays~\cite{opal-PN393} is also discussed here, due to the 
unavailability of the OPAL  speaker.

\section{Fermi-Dirac correlations in \boldmath{\PLL} system}

The FD correlations in \PLL\ system were studied using 
3.9~million hadronic \PZ\ decays recorded by the ALEPH detector
%from 1992 to 1995 
on and around the \PZ\ peak. 
A sample of  2133 pairs with $Q < 10$~GeV was obtained.

In the analysis, three reference samples were used:
A) simulated pairs from JETSET MC without FD correlations, where
\mbox{$C(Q) = P(Q)_{\rm data}/P(Q)_{\rm MC}$}; \\
B)
pairs obtained by event mixing, where
\mbox{$C(Q) = [P(Q)_{\rm data} / P(Q)_{\rm data}^\mathrm{mix}]/
              [P(Q)_{\rm MC}   / P(Q)_{\rm MC  }^\mathrm{mix}]$;}
C)
reweighted sample of mixed pairs, where
      \mbox{$C(Q) = P(Q)_{\rm data} / 
                    P(Q)_{\rm data,mix}^{\mathrm{reweighted}}$}.

\begin{figure}[htb]
  \vspace*{-12ex}
  \includegraphics[width=7.5cm]{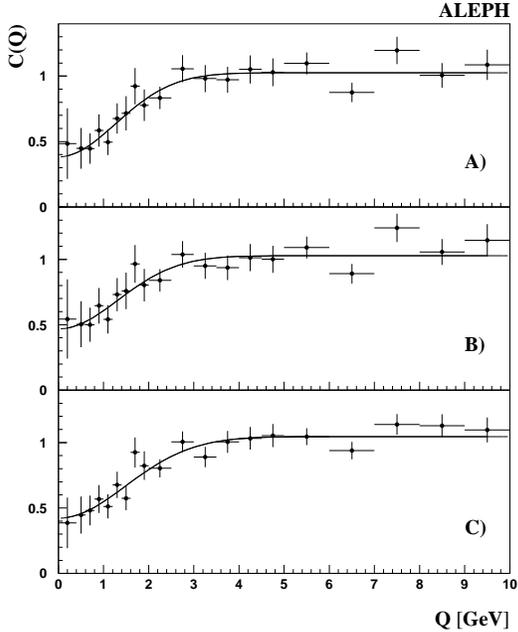} 
  \vspace*{-10ex}
  \caption{Correlation function for the \PLL\ pairs using reference 
           samples $A$, $B$, and $C$. The curves represent fits using 
           the parametrisation given in  Eq.~(\ref{eq:Goldhaber}).}
  \label{fig:CorrF}
  \vspace*{-4ex}
\end{figure}

\begin{figure}[tb]
  \vspace*{-2ex}
  \includegraphics[width=7.5cm]{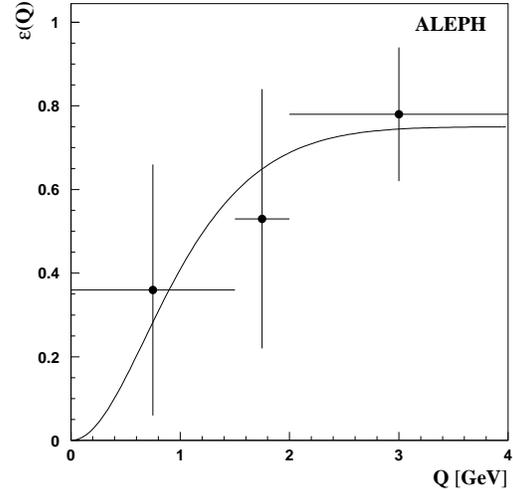} 
  \vspace*{-10ex}
  \caption{The fraction $\varepsilon(Q)$ of the $S=1$ contribution 
           for the \PLL\ data. The curve represents a fit using the 
           parametrisation  given in Eq.~(\ref{eq:epspar}).}
  \label{fig:epsilon}
  \vspace*{-4ex}
\end{figure}

The measured correlation functions are shown in Fig.~\ref{fig:CorrF},
parametrised with
\begin{equation}
C(Q)  = {\mathcal N} [1  + \beta  \exp( - R^2 Q^2)] \label{eq:Goldhaber}
\end{equation}
Consisted results are obtained for the three reference samples. 
The correlation function $C(Q)$ decreases for $Q < 2$~GeV; as $Q$ 
tends to zero, it approaches the value of  0.5, 
as expected for a statistical spin mixture ensemble. 
If this is interpreted as a FD effect and the parametrisation 
of Eq.~(\ref{eq:Goldhaber}) is used, the resulting values for the 
source size $R$ 
and for the suppression parameter $\beta$ are
\begin{xalignat*}{1}
R&=\phantom{-}0.11 \pm 0.02_{\rm stat} \pm 0.01_{\rm sys}\quad{\rm fm} \\
\beta &=    - 0.59 \pm 0.09_{\rm stat} \pm 0.04_{\rm sys}  
\end{xalignat*}
An alternative method to study the \PLL\ system is to measure the spin 
composition of the system using the angular distribution  
$\dr N/\dr y^*$, with $y^*$ the cosine of the angle between the two protons 
(antiprotons) in the di-hyperon centre-of-mass system. The measured 
$\dr N/\dr y^*$ distribution has contributions from both $S=1$ and $S=0$ 
states:
\begin{displaymath}
\frac{\dr N}{\dr y^*} = 
                 (1 - \varepsilon) \frac{\dr N}{\dr y^*}\bigg{|}_{S=0} + 
                      \varepsilon  \frac{\dr N}{\dr y^*}\bigg{|}_{S=1} 
\end{displaymath}
where $\varepsilon$  is the fraction of $S=1$ contribution. By fitting 
the $\dr N/\dr y^*$ distribution in different $Q$ ranges, the dependence
$\varepsilon(Q)$ is obtained. But $\varepsilon(Q)$ can also be defined 
as $\varepsilon(Q) =  C(Q)_{S=1} /[C(Q)_{S=0} + C(Q)_{S=1}]$, 
with $C(Q)_{S=1}$ and $C(Q)_{S=0}$ the contributions of $S=1$ and $S=0$
states to the correlation function $C(Q)$. Using the parametrisation 
given in  Eq.~(\ref{eq:Goldhaber}), one obtains for a statistical spin 
mixture ensemble 
\begin{equation} 
\D \varepsilon(Q)  = 0.75  \frac{1  -     \gamma \exp( - R^2 Q^2)}
                                {1  - 0.5 \gamma \exp( - R^2 Q^2)}  
                    \label{eq:epspar}
\end{equation}
with $\gamma=-2\beta$.
The distribution $\varepsilon(Q)$ is shown in Fig.~\ref{fig:epsilon},
fitted with the parametrisation given in Eq.~(\ref{eq:epspar}) 
with $\gamma$ fixed to one. 
The state $S=1$ dominates for $Q > 2$~GeV, but it is suppressed for 
$Q < 2$~GeV. The value of the source size estimated from 
$\varepsilon(Q)$ is $R = 0.14 \pm 0.09_{\rm stat} \pm 0.03_{\rm sys}$~fm, 
in agreement with the value obtained from the correlation function. From 
comparison with results from \Pgppm\Pgppm\ and \PKpm\PKpm\ correlation
measurements, one observes that the source dimension decreases with 
increasing mass of the emitted particle. 

\section{Colour reconnection in \PW-pair decays}

\begin{figure}[hbt]
  \vspace*{-5ex}
  \includegraphics[width=7.5cm]{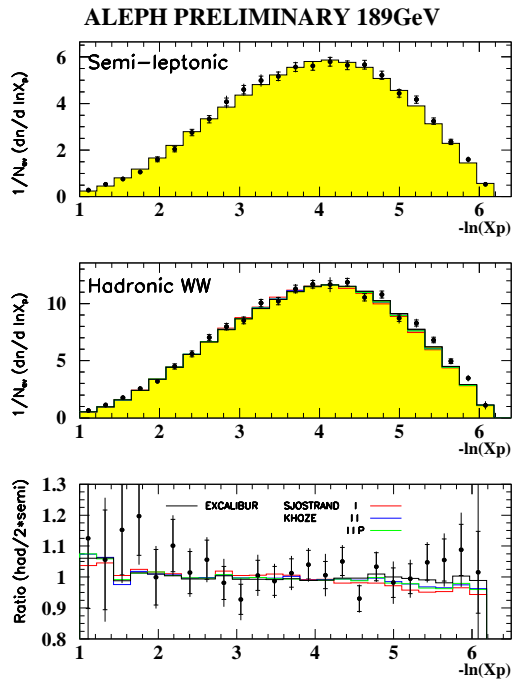} 
  \vspace*{-10ex}
  \caption{The $-\ln x_p$ distributions for semileptonic and hadronic 
           \PW-pair decays for data and MC models with and without CR.}
  \label{fig:CR}
  \vspace*{-3ex}
\end{figure}

The colour reconnection in $\Pep\Pem \to \PWp\PWm$ was studied in a 
data sample of 174.2 pb$^{-1}$ at a centre-of-mass energy of 
$\sqrt{s}=189$~GeV. Hadronic and semileptonic 
\PW-pair decays were selected and the 
experimental distribution $-\ln x_p$ of the charged particles was 
compared for each event type to the MC models KORALW and EXCALIBUR 
without CR and to EXCALIBUR with CR. 
Here $x_p=p/(\sqrt{s}/2)$ is the scaled momentum of a particle.
KORALW and EXCALIBUR are used to generate \PW-pairs; both
are coupled to JETSET for the hadronization part. Three CR 
models~\cite{aleph-CR}, denoted as $I$, $II$ and $IIP$ and 
implemented in JETSET, were compared to data. The distribution 
obtained for the data and the MC models are shown in Fig.~\ref{fig:CR}.
The ratio of the multiplicity in fully-hadronic events to twice the 
multiplicity in semileptonic events is also shown. 
The multiplicities within the experimental acceptance for the 
semileptonic channel are
$N_{ch}^{l\nu \Pq\Paq}=17.53 \pm 0.19 \pm 0.24$ for data and 
$N_{ch}^{l\nu \Pq\Paq}=17.41 \pm 0.04 \pm 0.29 $ for MC~(no CR), 
giving a difference between data and MC of $ 0.12 \pm 0.42$. 
For the fully hadronic channel, the multiplicity is   
$N_{ch}^{\Pq\Paq\Pq\Paq}=35.52 \pm 0.22 \pm 0.43 $ for data and 
$N_{ch}^{\Pq\Paq\Pq\Paq}=34.77 \pm 0.04 \pm 0.58 $ for MC~(no CR), 
which gives a difference of $0.75  \pm 0.76$. The difference
$N_{ch}^{\Pq\Paq\Pq\Paq} - 2N_{ch}^{l\nu \Pq\Paq}$ is 
$0.47 \pm 0.44 \pm 0.26 $ for data, $-0.05 \pm 0.09$ for MC (no CR)
and $0.52 \pm 0.52$ for the difference between data and MC. 
No colour reconnection was observed in the data, but at the current 
level of statistical precision both models with and without CR are 
compatible with the experimental results.

\section{Bose--Einstein correlations  in \PW-pair decays} 

The Bose--Einstein correlations in \PW-pair decays were studied using 
data recorded by the ALEPH detector at centre-of-mass energies of 
172, 183 and 189~GeV. 
For the tuning of the MC models of BE correlations, \PZ\ data recorded 
at 91.2~GeV with the same detector configuration as for \PWp\PWm\ 
events were used. Pairs of unlike-sign pions were chosen as reference 
sample and the correlation function was defined as
\begin{equation} 
\D    R^*(Q) = \frac{ \D  \left( \frac{ N_{\pi}^{++,--}(Q) }
                                      { N_{\pi}^{+-}   (Q) }  
                          \right)^\mathrm{data}                }
                    { \D  \left( \frac{ N_{\pi}^{++,--}(Q) }
                                  { N_{\pi}^{+-}   (Q) }  
                          \right)^\mathrm{MC}_\mathrm{no \;BE} } \;.
\label{eq:BEcor}
\end{equation} 
The correlation function was parametrised with
\begin{equation} 
R^*(Q) = \kappa (1+\epsilon Q)(1+\lambda \mathrm{e}^{-\sigma^{2}Q^{2}} )
\label{eq:BEpar}
\end{equation} 
where the term $1+\lambda \mathrm{e}^{-\sigma^{2}Q^{2}}$ describes the BE 
effect. The two parameters $\lambda$ and $\sigma$ characterise the 
effective strength of the correlations and the source size, 
respectively. The term $1+\epsilon Q$ takes into account some long 
range correlations due to the charge and the energy-momentum 
conservation, while $\kappa$ is a normalisation factor.

The BE correlations were first measured in \PZ\ decays. A MC 
simulation of the BE effect with the JETSET BE$_3$ 
model~\cite{JETSET-BE3} was then tuned on these data. The tuned 
parameters were $\lambda_\mathrm{input}=2.3$ and
$R_\mathrm{input}=0.26$~GeV. As \PW\ bosons do not decay 
into b quarks, the BE effect was determined separately in an udsc 
sample and in a b sample. Two b samples of different purities were
tagged~\cite{b-tagging} and the parameters $\lambda_\mathrm{b}$ and
$\lambda_\mathrm{udsc}$ were determined. Using these parameters, the 
\mbox{b$\bar{\mathrm{b}}$} component was replaced by an udsc 
component. The agreement between the MC model and the udsc data is 
good; residual discrepancies between them are corrected bin by bin.
 
\begin{figure}[htb]
  \vspace*{-3ex}
  \includegraphics[width=7.5cm]{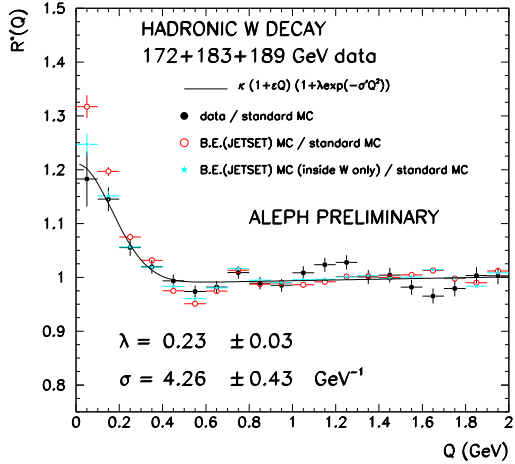} 
  \vspace*{-9ex}
  \caption{The correlation function $R^\star (Q)$ for fully hadronic 
           \PW-decays, compared to MC models of BE correlations.}
  \label{fig:BEhad}
  \vspace*{-4ex}
\end{figure}

The prediction of the MC model tuned and corrected on \PZ\ data was 
then checked on semileptonic \PW-pairs decays. They were found to be 
in very good agreement. For the hadronic \PW-decays, two cases were 
considered in the MC model: pions from different \PW's may exhibit
BE correlations (denoted as $BEB$) and only pions from the same \PW\
exhibit BE correlations (denoted BEI). The comparison between these
two MC models and data is shown in Fig.~\ref{fig:BEhad}. 
The result of the fit with the parametrisation given in 
Eq.~(\ref{eq:BEpar}) is also shown. All four parameters 
$\kappa$,  $\epsilon$, $\lambda$ and $\sigma$ were free
in this fit. The values of $\lambda$ and $\sigma$ were 
used to compute an integral of the BE signal
$I_{BE}=\int_0^\infty \lambda \mathrm{e}^{-\sigma^{2}Q^{2}}\dr Q 
          \propto \lambda / \sigma $ 
for data and MC.
A second fit with $\kappa$,  $\epsilon$ and $\sigma$ fixed
and  $\lambda$ free was also made to the first four bins
only, where the effect is expected to be maximum. The differences 
between the data and the MC models for $\lambda$ from the 
one-parameter fit are
\begin{xalignat*}{1}
\lambda^\mathrm{data}-\lambda^\mathrm{MC \; BEB}  = & 
                                           -0.088\pm0.026\pm0.020 \\
\lambda^\mathrm{data}-\lambda^\mathrm{MC \; BEI}  = &
                                           -0.019\pm0.026\pm0.016
\end{xalignat*}
while for the $I_{BE}$ quantity they are
\begin{xalignat*}{1}
\hspace*{-13pt}
I_{BE}^\mathrm{data}-I_{BE}^\mathrm{MC \; BEB} = &  
                                            -0.0217\pm0.0062\pm0.0048 \\
\hspace*{-13pt}
I_{BE}^\mathrm{data}-I_{BE}^\mathrm{MC \; BEI} = &
                                            -0.0040\pm0.0062\pm0.0036 
\end{xalignat*}
The first error is the statistical error, the second is the 
systematic one.  A better agreement is obtained for the JETSET model
with BE correlations present only for pions coming from the same
\PW\ boson. The JETSET model which allows for BE correlations
between pions from different \PW's is disfavoured by both $\lambda$
and $I_{BE}$ variables at $2.7\sigma$ level.  

\section{Bose--Einstein correlations  in \PW-pair decays at 
         OPAL%\protect\footnote{The OPAL results were included here due 
             %                  to the unavailability of the OPAL 
             %                  speaker.}
        }

\begin{figure}[htb]
  \vspace*{-5ex}
  \includegraphics[width=7.5cm]{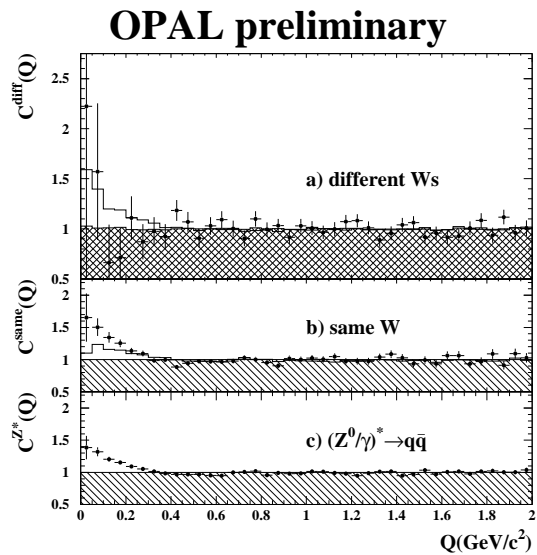} 
  \vspace*{-10ex}
  \caption{The correlation function for unfolded classes
           $C^{\mathrm{same}}(Q)$, $C^{\mathrm{diff}}(Q)$
           and $C^{\mathrm{Z^\star}}(Q)$.
          }
  \vspace*{-7ex}
  \label{fig:opalBE}
\end{figure}

The OPAL collaboration has analysed data recorded at centre-of-mass 
energies of 172, 183 and 189~GeV. Three mutually 
exclusive event sample were selected: the fully hadronic event 
sample $\PWp\PWm \rightarrow \Pq\Paq\Pq\Paq$, the semileptonic 
event sample  $\PWp\PWm \rightarrow l\nu   \Pq\Paq$ and the 
non-radiative $(\PZz/\Pgg)^{\star}$ event sample 
$(\PZz/\Pgg)^{\star} \rightarrow \Pq\Paq$. 
The correlation function  C(Q) was defined according to  
Eq.~(\ref{eq:BEcor}). For each sample, the correlation function was 
written as a combination of contributions from the various pure 
pion classes,  including the background. For hadronic event sample
one has
\begin{equation*} 
\begin{split}
C^{\mathrm{had}}(Q) =& 
        P^{\PW\PW}_{\mathrm{had}}(Q)  C^{\Pq\Paq\Pq\Paq}(Q)  + \\
   &[1-P^{\PW\PW}_{\mathrm{had}}(Q)]  C_{bg}^{\mathrm{Z^\star}}(Q) \;. 
\end{split}
\end{equation*} 
Similar expressions were written for $C^{\mathrm{semi}}(Q)$
and $C^{\mathrm{non-rad}}(Q)$ correlation functions. The probabilities 
$P^{\PW\PW}_{\mathrm{had}}(Q)$, etc. were taken from MC simulations 
without BE effect. 
Each correlation function $C^{\Pq\Paq\Pq\Paq}(Q)$, $C^{\Pq\Paq}(Q)$ and 
$C^{\mathrm{Z^\star}}(Q)$ was parametrised by
\begin{equation} 
C(Q)=N[1+f_\Pgp(Q)\lambda \exp(-Q^2R^2)] 
\label{eq:opalBEpar}         
\end{equation}
where $f_\Pgp(Q)$ is the probability of the pair to be a pair of pions.
A simultaneous fit was made to the experimental data, with the parameters
$N$, $\lambda$ and  $R$ free for each event class (nine free parameters).
All three classes exhibit BE correlations with consistent $R$ and 
$\lambda$ parameters. BE correlations were then investigated separately
for pions coming from the same \PW\ and from different \PW's. The 
correlation function for the hadronic event sample was written as 
\begin{eqnarray*} 
\lefteqn{C^{\mathrm{had}}(Q) =
          P^{\mathrm{same}}_{\mathrm{had}}(Q) C^{\mathrm{same}}(Q) +  
          P^{\mathrm{Z^\star}}_{\mathrm{had}}(Q)
                                       C_{bg}^{\mathrm{Z^\star}}(Q) 
        } \hspace*{7ex} \\
    && + \; [1-P^{\mathrm{same}}_{\mathrm{had}}(Q)-
            P^{\mathrm{Z^\star}}_{\mathrm{had}}(Q)] C^{\mathrm{diff}}(Q) 
\end{eqnarray*} 
where  $C^{\mathrm{same}}(Q)$, $C^{\mathrm{diff}}(Q)$ and  
$C_{bg}^{\mathrm{Z^\star}}(Q)$ are the correlation functions for pions 
from the same \PW, from different \PW's and from  
$(\PZz/\Pgg)^{\star} \rightarrow \Pq\Paq$ events. Similar expressions 
were written for $C^{\mathrm{semi}}(Q)$ and $C^{\mathrm{non-rad}}(Q)$. 
The correlation functions $C^{\mathrm{same}}(Q)$, $C^{\mathrm{diff}}(Q)$
and $C^{\mathrm{Z^\star}}(Q)$ were unfolded from the data and are 
shown in Fig.~\ref{fig:opalBE}. 
They were then parametrised by
Eq.~(\ref{eq:opalBEpar}) and simultaneous fits to the experimental 
distributions were performed. Three different cases were considered:
1) the same source-size $R$ for all event classes; 
2) different $R$ parameters  for each class, and 
3) $R^{\mathrm{diff}}$ is related to $R^{\mathrm{same}}$ using
the theoretical prediction 
$(R^{\mathrm{diff}})^2 = 
                      (R^{\mathrm{same}})^2+4\beta^2\gamma^2c^2\tau^2$.
The results obtained for the parameter $\lambda$ for the third case are
  \vspace*{-0.5ex}
\begin{alignat*}{2}
&\lambda^\mathrm{same}    & =  0.69 \pm 0.12 \pm 0.06 \\
&\lambda^\mathrm{diff}    & =  0.05 \pm 0.67 \pm 0.35 \\ 
&\lambda^\mathrm{Z^\star} & =  0.43 \pm 0.06 \pm 0.08
  \vspace*{-0.5ex}
\end{alignat*}
At the current level of statistical precision
it is not possible to determine if correlations between pions from 
different \PW's exist or not.

\section{Conclusion}

The ALEPH analyses of Fermi--Dirac correlations of \PLL\ pairs in 
hadronic \PZ\ decays and of colour reconnection and Bose--Einstein 
correlations in \PW-pairs decays have been presented.
A depletion of events are observed for the region $Q < 2$~GeV 
in the \PLL\ system.
In the analysis of \PW-pair decays, no colour reconnection effects are 
observed, but models which predict such effects can not be excluded.
The Bose--Einstein correlations measured in  \PW-pair decays are 
reproduced by a Monte Carlo model with independent fragmentation of 
the two \PW's, while a variant of the same model with  Bose--Einstein 
correlations between decay products of different \PW's is disfavoured 
at $2.7\sigma$. 
The OPAL analysis of Bose--Einstein correlations in \PW-pair decays 
has also been presented. At the current level of statistical precision
it was not possible to determine if correlations between pions from 
different \PW's exist or not.

{\bf A. Simon}, University of Freiburg (Germany) \\
{\it
The experiments obviously use different methods to study BE
correlations. It is technically possible to agree on one dataset and 
check the various methods on their systematics?
}

Answer:
{\it
Each experiment has optimised the analysis on detector characteristics and
on the aspects of BE correlations which were considered important to be
studied with the available statistics. The methods are therefore
different; this is a normal situation, there is no need to have
coordinated analyses until each experiment obtains a set of ``final
results''. A too early coordination and convergence of the methods can
reduce the quality of the analyses. Once the experiments obtains 
``final results'', an intercomparison is meaningful, followed by cross 
checks of the methods used by the other experiments (within the limits of
statistics and... manpower). In fact, a LEP group was formed recently 
to understand the differences between the results of the four LEP experiments. 
}

%%%%%%%%%%%%%%%%%%%%%%%%%%%%%%%%%%%%%%%%%%%%%%%%%%%%%%%%%%%%%%%%%%%%%
\end{document}